\documentclass[preprint,12pt]{elsarticle}
\usepackage[top=3cm,bottom=3cm,left=3cm,right=3cm]{geometry}
\usepackage[cmex10]{amsmath}
\usepackage{comment}
\usepackage{ifpdf} 
\usepackage{graphicx} 
\usepackage{epstopdf} \usepackage[font=small,labelfont=rm]{caption}
\usepackage[utf8x]{inputenc}
\usepackage{tabularx,ragged2e}
\usepackage{algorithm}
\usepackage{hyperref}
\usepackage{algorithmicx}
\usepackage{siunitx}
\usepackage{algpseudocode}
\usepackage{graphics}
\usepackage{etoolbox}
\usepackage{nomencl}
\makenomenclature
\usepackage{amssymb}
\usepackage{lineno}
\usepackage[dvipsnames]{xcolor}

\begin{document}

\begin{frontmatter}

\title{Value of Optimal Trip and Charging Scheduling of Commercial Electric Vehicle Fleets with Vehicle-to-Grid in Future Low Inertia Systems}
\author[1]{Alicia Blatiak\corref{cor1}}
\ead{alicia.blatiak@imperial.ac.uk}
\author[1]{Federica Bellizio}
\ead{f.bellizio18@imperial.ac.uk}
\author[1]{Luis Badesa}
\ead{luis.badesa@imperial.ac.uk}
\author[1]{Goran Strbac}
\ead{g.strbac@imperial.ac.uk}

\cortext[cor1]{Corresponding author}
\address[1]{Imperial College London, South Kensington, London SW7 2BU, United Kingdom}

\begin{abstract}
The electrification of transport is seen as an important step in the global decarbonisation agenda. With such a large expected load on the power system from electric vehicles (EVs), it is important to coordinate charging in order to balance the supply and demand for electricity. Bidirectional charging, enabled through Vehicle-to-Grid (V2G) technology, will unlock significant storage capacity from stationary EVs that are plugged in. To take this concept a step further, this paper quantifies the potential revenues to be gained by a commercial EV fleet operator from simultaneously scheduling its trips on a day-ahead basis, as well as its charging. This allows the fleet to complete its trips (with user defined trip length and distance), while taking advantage of fluctuating energy and ancillary services prices. A mathematical framework for optimal trip scheduling is proposed, formulated as a mixed-integer linear program, and is applied to several relevant scenarios of the present and future British electricity system. It is demonstrated that an optimal journey start time can increase the revenue of commercial fleets by up to 38\% in summer and 12\% in winter. This means a single EV from the maintenance fleet can make additional annual revenue of up to £729. Flexible trip schedules are more valuable in the summer because keeping EVs plugged in during peak solar output will benefit the grid and the fleet operators the most. It was also found that a fleet of 5,000 EVs would result in the equivalent $CO_2$ of removing one Combined Cycle Gas Turbine from the system. This significant increase in revenue and carbon savings show this approach is worth investigating for potential future application.

\end{abstract}

% \begin{highlights}
%      \item Quantification of the value of optimal trip scheduling considering battery degradation and frequency response prices 
%      \item Characterisation of flexibility from commercial EV fleets using real world data
%      \item  Revenue increases of commercial fleets by up to 38\% in summer and 12\% in winter using optimal journey start times
%      \item Large reductions of $CO_2$ emissions equivalent of one Combined Cycle Gas Turbine using a fleet of 5,000 EVs
% \end{highlights}


\end{frontmatter}

\vspace*{-6mm}
\vspace*{0.5mm}
\section*{\vspace*{0.5mm}Nomenclature}
\begin{tabular}{p{12mm} l}
%\vspace*{8.5mm}\\

\hspace{-3mm} $\beta$ & Objective function in Equation 1\\
\hspace{-3mm} $\delta$ & Penalty term\\
\hspace{-3mm} $\eta_{c}$ & Charge efficiency\\
\hspace{-3mm} $\textrm{CCGT}$ & Combined Cycle Gas Turbine\\
\hspace{-3mm} $\eta_{c}$ & Discharge efficiency\\
\hspace{-3mm} $\pi_t^{b}$ & Hourly import energy (buy) price (£)\\
\hspace{-3mm} $\pi^{BSup}$ & Hourly upward balancing service price (£)\\
\hspace{-3mm} $\pi_t^{s}$ & Hourly export energy (sell) price (£)\\
\hspace{-3mm} $BSup_{t}^{c}$ & Upward balancing services\\
\hspace{-3mm} $BSup_{t}^{d}$ & Downward balancing services\\
\hspace{-3mm} $C_t$ & Charging rate at time step $t$ (kW)\\
\hspace{-3mm} $D_t$ & Discharging rate at time step $t$ (kW)\\
\hspace{-3mm} $\textrm{DC}$ & Dynamic Containment\\
\hspace{-3mm} $\textrm{DC}_\textrm{EVs}$ & Dynamic Containment provision from Electric Vehicles\\
\hspace{-3mm} $E_{max}$ & Maximum battery capacity (kWh)\\
\hspace{-3mm} $E_{min}$ & Minimum battery capacity (kWh)\\
\hspace{-3mm} $E_{start}$ & Battery energy at start of time period $T$ (kWh)\\
\hspace{-3mm} $E_{req}$ & Battery energy required at end of time period $T$ (kWh)\\
\hspace{-3mm} $E_{t}$ & Battery energy at time step $t$ (kWh)\\
\hspace{-3mm} $\textrm{EV}$ & Electric Vehicle\\
\hspace{-3mm} $\textrm{E-VRP}$ & Electric Vehicle Routing Problem\\
\hspace{-3mm} $E_{tr}$ & Energy needed for travel (kWh)\\
\hspace{-3mm} $f_0$ & Nominal frequency of the power grid (Hz)\\
\hspace{-3mm} $\Delta f_{max}$ & Lowest nadir admissible to avoid Under-Frequency Load Shedding\\
\hspace{-3mm} $\textrm{FR}$ & Frequency Response ($\textrm{MW}$)\\
\hspace{-3mm} $\textrm{GB}$ & Great Britain\\
\hspace{-3mm} $\textrm{H}$ & Total system inertia ($\textrm{MVA}\cdot \textrm{s}$)\\
\hspace{-3mm} $\textrm{H}_\textrm{CCGTs}$ & Inertia from Combined Cycle Gas Turbines ($\textrm{MVA}\cdot \textrm{s}$)\\
\hspace{-3mm} $\textrm{H}_\textrm{floor}$ & Floor for inertia ($\textrm{MVA}\cdot \textrm{s}$)\\
\hspace{-3mm} $P\textsubscript{infeed}$ & Largest power infeed in the national system\\
\hspace{-3mm} $P^{max}$ & Maximum power rating of charger (kW)\\
\hspace{-3mm} $\textrm{PFR}$ & Primary Frequency Response ($\textrm{MW}$)\\
\hspace{-3mm} $\textrm{RES}$ & Renewable Energy Sources\\
\hspace{-3mm} $t$ & Time step\\
\hspace{-3mm} $t_s$ & Minimum time required to sustain balancing service\\
\hspace{-3mm} $t_{start}^{*}$ & Optimal trip start time\\
\hspace{-3mm} $T$ & Total time period\\
\hspace{-3mm} $T_{grid}$ & Time period that electric vehicle is connected to the grid\\
\hspace{-3mm} $T_{window}$ & Time period where trips are permissible\\
\hspace{-3mm} $\textrm{V2G}$ & Vehicle-to-Grid\\
\hspace{-3mm} $\textrm{VRP}$ & Vehicle Routing Problem\\
\hspace{-3mm} $x_1$ & Auxiliary decision variable\\
\hspace{-3mm} $y$ & Binary variable\\

\end{tabular}

\section{Introduction}

With the global agenda to reduce reliance on fossil fuels and reduce pollutants and carbon emissions, the transition from internal combustion engine vehicles to electric vehicles (EVs) is seen as an important one. In $2020$, despite the Covid-$19$ pandemic, a record three million new EVs were registered globally, a $41\%$ increase from the previous year. By $2030$, the global EV fleet is projected to reach $145$ million, rising to potentially as much as $230$ million, if governments accelerate efforts to reach climate goals \cite{InternationalEnergyAgency2021Global2021}. The added charging loads associated with an increasing penetration of EVs, could pose a threat to the stability and reliability of the power system \cite{Bozic2015ImpactReliability}.

However, there is also an opportunity to better integrate the transport and power systems, with the extra storage capacity offered by the EV fleet to the system operator. Storage is a type of flexibility, or adjustment of generation or consumption, which will be needed to ensure the future low-carbon power system can maintain secure operation. By accessing more flexibility, system operators can operate the system very close to its limits and reduce the safety margins. This would result in an improved utilization of the existing grid assets and in a reduction of the investment costs as no redundant equipment would be anymore needed to maintain the system within adequate security levels. Therefore, carbon emissions targets can be achieved while making significant savings by reducing the need for investment in low-carbon generation, reducing system operating costs and reducing network reinforcement costs \cite{ImperialCollegeLondon2017RoadmapChange}. During the Covid-$19$ pandemic, a glimpse of this future power system was seen, due to the lower demand and high renewable generation seen in May to July, $2020$, in Great Britain. Operating costs in the form of ancillary services increased by \textsterling$200$m compared to the same period in $2019$. The study in \cite{Badesa2021AncillaryFuture} also showed that in the $2030$ future power system, these ancillary service costs could reach $35\%$ of total operating costs.

\subsection{Existing Approaches}

Electric vehicles research exists in many fields and from many perspectives. Regarding their interaction with the electricity network, there are two established fields that investigate this issue, power systems and transport studies. The latest developments in these fields will be summarised in this section, describing the state of the art. There is one emerging field where these disciplines are beginning to overlap, and where this paper contributes. It is important that both perspectives, power systems and transport studies (which can be generally summarised as the charging scheduling of the EVs and the travel scheduling of the EVs), are both considered as transport is electrified. This is due to the increasingly relevant and complex interaction between energy and transport systems moving forward.

There has been significant research in the field of modern power systems, investigating how to take advantage of the flexibility EVs have to offer. Often, this is from a power system perspective, where EV charging scheduling, i.e.~optimising charging or discharging schedules according to either energy arbitrage or balancing service provision, has been a focus. Early work in this area was developed in \cite{Kempton2005a}, which calculated the potential capacity and net revenues for EVs supporting the grid with Vehicle-to-Grid (V2G) technology. The potential for EVs to provide grid services such as peak power, spinning reserves are regulation, was outlined and the societal benefits also discussed. An optimal bidding strategy of EVs participating in electricity markets was formulated in \cite{Vagropoulos2013}, where an aggregator of EVs bids strategically, considering uncertainties in price and driving behaviour. Another operational tool was developed in \cite{Sortomme2012OptimalServices}, allowing the aggregator to optimise energy and ancillary services scheduling together. This allowed for higher profits for aggregators, as well as lower costs of EV charging for the customers. EV charging is optimised to minimise carbon emissions and wind curtailment in  \cite{Dixon2020SchedulingCurtailment}, showing EVs can absorb excess wind generation in GB. More recently, EVs have also been studied as a potential source of synthetic inertia for low inertia power systems \cite{Magdy2021AGrids} and of reactive power \cite{Mojdehi2016TechnicalEVs}. A technical review on approaches for EV demand management is provided in \cite{Teng2020TechnicalIntegration} and covers multiple electric power markets. 

A recent review of progress in utilizing EVs for ancillary services is presented in \cite{Ravi2022UtilizationPerspectives} and is done so from several perspectives, including state of the art technology and control methods. An extensive review of charging and discharging strategies, including various objective functions for optimising EVs in the power system, has been carried out in \cite{El-bayeh2021ChargingSurvey}. A common objective function identified in the literature is maximising revenue for the EV owners, fleet operators or aggregators participating in the energy and balancing service markets. In terms of physical constraints and market participation, the basis of modelling EVs as storage is similar to energy storage systems, for example in \cite{Rodrigues2018Risk-averseStorage}. It is worth noting that in most power system related studies, the driving behaviour of the EVs is not a focus, rather a predetermined input and often fits a Gaussian distribution based on trends in driving pattern data (for example, in \cite{Vagropoulos2013} and \cite{Xu2017}). Balancing (or ancillary) services, including frequency response (FR), have also been of particular interest because of their market value \cite{Kempton2005a} and the minimal stress on EV batteries (i.e.~when providing availability for the majority of the time, thereby reducing the charge cycling). Utility fleets, with their ease of aggregation by fleet operators or aggregators, can provide the minimum bidding capacity in order to participate in the frequency response market, currently set to $1$MW. These fleets are also highly predictable, allowing for a greater certainty that the capacity offered will be met, compared to, for example, aggregating private vehicle fleets \cite{Dixon2020ElectricNetworks}.

A limitation of these studies of EV charging and discharging schedules is that most do not consider vehicle trip timings or routes as decision variables in their approaches (i.e.~driving patterns are either fixed inputs or the uncertainty in plug-in times is accounted for and modelled). There is another field that focuses on these issues but without the power system dimension. The Vehicle Routing Problem (VRP), specific to EVs (E-VRP), approaches the optimal operation of EVs from the driving pattern and infrastructure perspective. Here, the objective function is often to plan the optimal route, as opposed to the optimal charging schedule, as often seen in power systems research. This optimal route takes into account traffic, charging needs and other constraints \cite{Erdelic2019AApproaches}. A limitation of these works is that although the prices of the use of charge-points are considered, these are not clearly linked to ancillary service market price. The prices are described as being related to electricity `cost', service and parking fees \cite{Wang2018OptimisingObjectives}. Recent work in \cite{Tookanlou2021AOperation} has integrated the optimal charging schedule with a vehicle routing problem, to produce maximum revenues for various agents, while also solving which route EVs should take. In \cite{Trivino-Cabrera2019JointV2G}, a mixed integer linear program was formulated to create charging schedules for EVs, as well as an optimal route, which considered distribution network constraints. The approach taken in \cite{Liu2020JointNetwork} also involved charging schedule optimisation and routing the EVs under time delay tolerance. Time delay tolerance is related to how long an EV is willing to wait in a queue to charge. A limitation of these works is that they do not consider the perspective of commercial fleet operation, nor do they account for ancillary service participation or prices. So in other words, the element of tapping into the flexibility EVs have to offer the power system is not fully explored in combination with the optimisation of the trip itself. In order to effectively integrate EVs, making them useful to the system operator, as well as maximise their own revenues, it is important that the ancillary service market is considered.

The provision of ancillary services, which is significantly higher with bi-directional charging (i.e.~V2G technology), has not been identified in the E-VRP literature, nor is the provision of energy arbitrage. This means that optimising EV operations is either done from the power system operation perspective, as per the power system research field, or from the EV operation perspective, as seen in the Transport Studies field. Clearly, with the electrification of transport, the power and transport systems will be much more interconnected and inter-dependent and there is currently a significant gap in this area of research. It is this precise limitation in the current literature that the present work seeks to address.

There is an emerging area of research that investigates how to coordinate charging EVs considering their transport services, too. To date, the use case for this approach is a shared EV fleet or an EV taxi fleet, no work has been done on the commercial maintenance or delivery fleet use case. In \cite{Zhao2021EVSystems}, historical taxi travel data and electricity prices were used to demonstrate how to maximise profits for the EV fleet owner. It was found that there was a higher demand for passenger taxi trips in the afternoon, when electricity prices were low. During the evening peak electricity prices, it was profitable for EVs to be plugged in as they could maximise profits by discharging and selling their energy. Although the work highlights the synergies between transport demand and electricity price, ancillary service provision is not considered. Also, energy arbitrage increases battery cycling and hence degradation more than ancillary service provision. In a future low-inertia system, frequency response will become increasingly important. Ancillary service provision from EVs could reduce system costs in a future GB electricity grid by up to \textsterling12,000 a year and reduce \textrm{$CO_2$} emissions by 60 tonnes per year \cite{Malley2020ValueServices}. The authors in \cite{Al-obaidi2021AdaptivePricing} designed a central controller for an EV taxi fleet that does include scheduling charging and discharging for ancillary service provision. The work does not schedule trips themselves, though, as this is determined by the taxi passengers. In the present work, scheduling is carried out both for trips and for charging to provide ancillary services, considering transport energy requirements and trip lengths. In summary, the main novelty of this paper compared to previous approaches is the co-optimisation of trip schedules alongside energy and ancillary services for commercial EV fleets.

\subsection{Motivation and Contributions}

The motivation of this work is to further explore and evidence the opportunities to integrate the energy and transport systems, in order to quantify the benefits to the future power system and to EV fleet operators. It is clear that there is value in optimising charging EVs to access the flexibility, in terms of energy, they have to offer. There also exists a body of work evidencing the value in the optimising the journeys of EVs, considering the changes to the transport system (including driving behaviour) that shifting to battery powered vehicles will have and mitigating this effect. In this work, the authors seek to quantify the value of combining these approaches and to answer the question, what will the value of simultaneously scheduling both EV charging and trips be in the future power system? This is particularly interesting considering the increased penetration of renewable energy in the future and the effect on ancillary service and energy prices. Frequency response is the ancillary service modelled here, as it is of major concern for a low-inertia grid with a high penetration of renewable generation. The generation mix will radically change the time of day that ancillary services, and hence EV flexibility and availability, will be useful. As a result, the possibility of scheduling trips, in addition to charging, to account for this is what motivates this work. 

As far as the authors are aware, no work has sought to quantify the value of optimal trip and charging schedules of commercial EV fleets participating in ancillary service markets. This is important given the projected uptake of EVs and the opportunity they present to provide significant flexibility and reduce power system operational costs, along with other distributed energy resources \cite{ImperialCollegeLondon2017RoadmapChange}. This paper seeks to consider both the fleet operation and the provision of ancillary services to the grid by optimising revenue for the commercial fleet owner considering energy and ancillary service prices. It does this by extending current approaches in optimising EV charging schedules to assess the value in scheduling trips, as well as incorporating both future prices / scenarios. This is valuable because this approach allows the fleet owners to benefit by increasing their revenue, as well as the power system benefiting from an increased capacity in ancillary services. As discussed in \cite{Badesa2021AncillaryFuture}, the operational costs from ancillary services is likely to increase in future power system scenarios, where more sources of renewable energy generation are present. This is an opportunity to utilise the new sources of flexibility the EVs present, while fulfilling the requirements of the EV fleet to perform its operations. Specifically, this paper considers the operations of the commercial fleet, where trips lengths and distances (i.e.~energy required for travel) are known, day-ahead user inputs. This work is different to existing work in that it focuses on commercial (e.g.~delivery and maintenance) fleets with known transportation needs and it considers the current and future power system scenarios to demonstrate the business case for commercial EV adoption. The real world commercial data obtained has not been seen elsewhere in the literature, and given its limited volume, data was synthetically generated for the case studies, based on distributions. The key contributions of this work can be summarised as follows:
\begin{itemize}
    \item Novel scheduling approach for EV fleet operators that includes both charging and trip times, generating additional revenues and exploiting synergies between the power system and transport system
    \item Quantification of the value of scheduling optimal journey times accounting for frequency response prices, limiting battery degradation and accurately capturing provision both from reducing charging and increasing discharging 
    \item First time accurate characterisation of flexibility of commercial EV fleets participating in ancillary service markets, inferred from real world data from a V2G demonstration project in the UK
    \item Comprehensive assessment of revenues of commercial V2G fleets considering present and future GB scenarios with various sensitivities, including societal benefits in terms of carbon emissions avoided 
    
    % previous contributions written
    
    % \item Individual EV day-ahead schedules - including optimal charging and trips times - are produced in order to maximise fleet operator revenue from frequency response market participation, while maintaining trip energy and duration requirements
    % \item Future frequency response and energy prices are simulated using future scenarios to quantify the potential value of commercial EV fleet participation % or to understand / predict the value of optimal scheduling of this sort in the future, which could be considered in the business models of current commercial fleets as they electrify? as a secondary revenue stream? 
    % \item Carbon emission reduction and Combined Cycle Gas Turbines (CCGTs) avoided from fleet participation in these markets are calculated for future scenarios
    % \item Real world commercial EV fleet data from Innovate UK V2G Demonstrator project, E-Flex, is used to generate synthetic data for $200$ EVs over one year
\end{itemize}

It should be noted that in this work, the EVs do not perform energy arbitrage, due to the adverse impacts on the battery by increased cycle \cite{Bishop2013EvaluatingEV}. 
%EVs with V2G compete most strongly with centralized generators when the service includes an availability (or capacity) payment and an added utilisation (or energy) payment. 
It has been found that EVs are only competitive on the electricity market, performing energy arbitrage, when the peak prices are unusually high \cite{Kempton2005a}. So although the fleet will optimise its charging schedule in order to charge when the price is cheapest, it will not discharge at peak price in order to generate more revenues by selling energy. This is achieved by assuming a zero sell price for energy.

The remainder of this this paper is structured as follows; Section \ref{methodology} provides the mathematical formulation of the problem, including the charging optimisation in Section \ref{optimal_ch}, as well as the trip optimisation in Section \ref{optimal_trip}. Case studies are described in Section \ref{case studies}, with the fleet data used detailed in Section \ref{fleet data}, scenarios and prices in Section \ref{scenarios and prices} and the results and discussion in Section \ref{results and discussion}. Finally, conclusions and further work are discussed in Section~\ref{conclusion}.  

\section{Methodology}\label{methodology}

\subsection{Charging Schedule Optimisation} \label{optimal_ch}

%The optimisation maximises revenues for the fleet operator by creating the best possible schedule for the following day. Considering energy and frequency response prices, as well as trip requirements, the model outputs a charging and trip schedule. % NOTE: can't aggregate AND optimise trips - prepare answer for this or mention in future work section)

The schedule - including charging and trips - for a commercial EV fleet is optimised over a day-ahead 24-hour horizon of time steps $t$ in total time period $T$. The EV can charge ($C_{t}$) and discharge ($D_{t}$), acquiring energy needed for travel ($E_{tr}$), as well as providing an upward balancing service by reducing charging ($BSup_{t}^{c}$) or increasing discharging ($BSup_{t}^{d}$) in the form of availability. The charging / discharging power of an EV can be continuously regulated between 0 and a maximum power level allowed by the charger ($P^{max}$), and they need to fulfill a predetermined energy requirement for the scheduled journey within the interval $T$. The trips that the EVs of the commercial fleet need to take, including the length of the trip in hours and the energy required (i.e.~the distance) is known ahead of time. This ensures the user experience is not affected, i.e. the fleet operator will input the trip distance and length for the EV, according to the work schedule, and the optimal start time for this trip as well as the charging schedule will be determined. The energy prices and the balancing service prices are also known ahead of the one-day time horizon, which is in line with the current wholesale energy and balancing service market arrangements.

The objective function maximises revenues from balancing services considering charging costs, while meeting the EV trip requirement for energy for travel, over the time periods that the EV is connected to the grid ($T_{grid}$). Smart charging takes into account the hourly import energy (buy) price $\pi_t^{b}$ and export (sell) price $\pi_t^{s}$.
Revenues from balancing services are from the availability of upward balancing service ($BSup_{t}^{c}$ and $BSup_{t}^{d}$) multiplied by the price ($\pi^{BSup}$), as shown in \eqref{eq:obj}. 

\begin{equation} \label{eq:obj} 
    \text{maximise} \left (\sum_{t\epsilon T^{grid}}  (BSup_{t}^{c}+BSup_{t}^{d}) \cdot \pi_t^{BSup} - C_{t} \cdot \pi_t^{b} + D_{t} \cdot \pi_t^{s} \right)
\end{equation}

The operating model of the EV includes several constraints. Constraint \eqref{eq:ev0} corresponds to the EV battery's energy balance, taking into account the energy needed for travel purposes as well as the losses caused by charging and discharging efficiencies.

\begin{equation} \label{eq:ev0} 
    E_{t} = E_{t-1} + C_{t} \Delta t \eta_{c} - \frac{ D_{t} \Delta t}{\eta_{d}} - {E_{tr}}, \forall t \in T
\end{equation} 

Each EV is assumed to depart from its grid connection point up to once within the time horizon and subsequently arrive back to its grid connection point only once during the same horizon. This is in line with the trend clearly seen in the real world data, whereby the commercial vehicles follow a schedule of one extended trip from the depot in one day. Outside of this single trip, the EVs are otherwise assumed to be connected to the grid, the hours of which constitute the time period $T_{grid}$. When the EVs are not connected to the grid, the charging, discharging and balancing services is zero, as per constraint \eqref{eq:ev1}.

\begin{equation} \label{eq:ev1} 
    C_{t} = D_{t} = BSup_{t}^{c} = BSup_{t}^{d} = 0, \forall t \notin T_{grid}
\end{equation} 

Constraint \eqref{eq:ev2} expresses the lower ($E_{min}$) and upper ($E_{max}$) bounds of the battery’s energy content, assumed to be the same for each EV.
\begin{equation} \label{eq:ev2} 
    E_{min} \leq E_{t} \leq E_{max}, \forall t \in T
\end{equation}

Constraint \eqref{eq:ev3} ensures that the energy level of the EV at the start of the trip ($E_{t_{start}}$) is at the required level ($E_{req}$) set by the fleet operator, in order to ensure the EV can carry out its work trip.
\begin{equation} \label{eq:ev3} 
    E_{t_{start}} = E_{req} 
\end{equation}

Constraints \eqref{eq:ev4} and \eqref{eq:ev5} represent the limits of the battery’s charging / discharging power, which depend on its power capacity $P^{max}$ and on the availability of the EV to be scheduled (i.e.~time period connected to the grid, $T_{grid}$). The binary variable $y$ ensures that charging and discharging do not happen simultaneously.
\begin{equation} \label{eq:ev4} 
    C_{t} \leq y\cdot P^{max}, \forall t \in T_{grid} 
\end{equation}
\begin{equation} \label{eq:ev5} 
    D_{t} \leq (1-y)\cdot P^{max}, \forall t \in T_{grid}
\end{equation}

Constraints (\ref{eq:pw1} - \ref{eq:pw3}) ensure that the balancing service availability committed takes into account the maximum power of the charger and whether the EV is charging or discharging. 

\begin{equation}\label{eq:pw1} 
     BSup_{t}^{d} + D_t \leq P^{max}, \forall t \in T_{grid}
\end{equation}

\begin{equation}\label{eq:pw2} 
     BSup_{t}^{c} \leq C_{t}, \forall t \in T_{grid}
\end{equation}

\begin{equation}\label{eq:pw3} 
     (D_{t} - C_{t}) + (BSup_{t}^{c} + BSup_{t}^{d}) \leq P^{max}, \forall t \in T_{grid}
\end{equation}

When committing balancing service availability, it is important to ensure that the energy is available for the service to be sustained for the minimum required time ($t_s$) if this service was called on. This applies to the balancing service from discharging ($BSup_{t}^{d}$) in particular, due to this being an upward balancing service. This is captured in constraints (\ref{eq:ev9}) - (\ref{eq:x1_2}) by considering the state of charge of the battery at the beginning and at the end of the time step. The balancing service provision from EVs only considers availability, rather than utilisation, which is assumed to occur infrequently \cite{Aunedi2020Whole-systemFleets}.
%Marko makes this assumption outright, might be a better ref for it

\begin{equation}
    E_{min} \leq \textrm{min}(E_{t}, E_{t-1}) + C_{t}\cdot \eta_{c} - \frac{ (D_{t} + BSup^{d}_{t})\cdot t_s} {\eta_{d}}, \forall t \in T_{grid}
\end{equation}

The above constraint can be linearised by introducing an auxiliary decision variable `$x_1$' and two additional constraints:
\begin{equation}\label{eq:ev9} 
    E_{min} \leq x_{1} + C_{t}\cdot \eta_{c} - \frac{ (D_{t} + BSup^{d}_{t})\cdot t_s} {\eta_{d}}, \forall t \in T_{grid}
\end{equation}

\begin{equation}\label{eq:x1_1} 
    x_1 \leq E_t
\end{equation}

\begin{equation}\label{eq:x1_2} 
    x_1 \leq E_{t-1}
\end{equation}

\subsection{Trip Schedule Optimisation} \label{optimal_trip}

The optimal trip start time ($t_{start}^{*}$) is found by maximising the revenue from the EV, according to the optimal charging schedule in Section {\ref{optimal_ch}}, across different feasible start times $t_{start}$ in which the EV could make the trip required. The variable $t_{start}$ informs when the EV is connected to the grid ($T_{grid}$ in \eqref{eq:obj}) and able to charge, discharge or provide frequency response. Therefore, a different optimal revenue corresponds to each $t_{start}$. To make the notation easier, the objective function in Eq.~\eqref{eq:obj} is defined as $\beta(t_{start})$.

The time that the trip is able to start is constrained by how long the trip is ($T_{trip}$), as well as the permissible travel window ($T_{window}$). For instance, if the 24-hour period starts at 7am with a trip required lasting five hours and a window of 15 hours (i.e.~the EV can be away from the charger between 7am and 9pm), the start time could be any hour from 7am to 4pm. The trip schedule optimisation problem is formulated as follows:
\begin{equation}
\begin{aligned}
& \underset{t_{start}}{\text{maximise}}
& & \beta(t_{start}) \\
& \text{subject to}
& & t_{start} < {T_{window}} - {T_{trip}}\\
& & & T_{grid} = T-(t_{start}+T_{trip})
\end{aligned}
\label{eq:trip}
\end{equation}

It is important to note that although the assumption is the trip length will remain the same regardless of the time of day it is taken, there is a level of uncertainty related to external factors, such as road blockages and traffic levels. Although in practice the fleet operator needs to create a schedule for the next day (hence the decision to formulate the scheduling problem here as such), there may be some flexibility by one or two hours on the day, to account for these uncertainties. For example, if there is a road blockage and the trip could be delayed by one or two hours. To investigate the effects of these external factors, a sensitivity analysis has been performed in Section \ref{SensitivityTraffic}.

\begin{figure}[h]
\centering
\includegraphics[width=0.9\textwidth, trim =0 0 0 0, ]{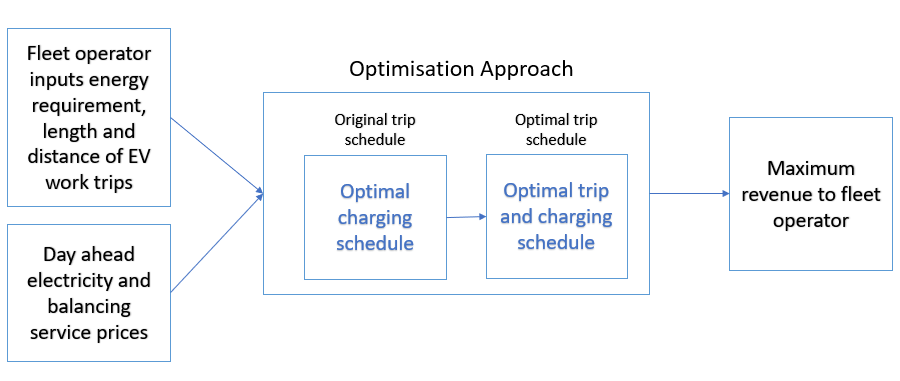}
\caption{Summary of workflow.}
\label{fig:workflow}
\end{figure}

Figure \ref{fig:workflow} shows an overview of the workflow of the above methodology, producing a quantification of the value from the optimal trip and scheduling approach. Table \ref{tab:vars} lists the input and decision variables used, including from the fleet operator and from the market.

\begin{table}[h]
\caption{Summary of variables used in optimisation approach.}
\centering
\begin{tabularx}{.8\columnwidth}{c| *{2}{c}}
\hline
 & \textbf{Input Parameters} & \textbf{Decision Variables} \\
\hline
Fleet Operator & $E_{tr}$, $E_{req}$, $T_{trip}$ & $t_{start}$, $C_t$, $D_t$ \\
Market Information & $\pi_t^{BSup}$, $\pi_t^b$, $\pi_t^s$ & $BSup_t^{c}$, $BSup_t^{d}$\\

\hline
\end{tabularx}
\label{tab:vars}
\end{table}

\subsection{EV Operation and Scenario Specific Constraints}{\label{scenarios and prices}}

In order to assess the value of the approach presented in this paper, different scenarios representing different modes of operation are modelled. This allows for insights into the various business cases for commercial EV adoption. EVs can operate in various modes, including dumb charging, smart charging and V2G. Here, the scenarios applied are described, along with a description of the state of pricing and price assumptions. In all scenarios, a zero sell price for energy is assumed, so although the EVs can discharge to the grid, energy arbitrage is not profitable. This is because the focus of this paper is the provision of upward balancing services from commercial EV fleets and energy arbitrage will potential degrade the battery by increasing the number of charge cycles \cite{Bishop2013EvaluatingEV}. % Goran said not to include arbitrage - on a 40kWh battery with efficiencies etc, it may not make sense, especially if there isn't a sell price in modern tariffs ?

%%%%%%%%%%%%%%%%%%%% scenario 1 - dumb charging %%%%%%%%%%%%%%%%%%%%%%
The first scenario, which is the baseline, is the `dumb charging' scenario. Here, EVs plug in and charge back up to their desired value (in this case 95\% SOC) immediately, regardless of the price of energy on return. Also, the scenario specific constraints include \eqref{eq:ev10} and \eqref{eq:ev11}, shown below. This modification to original constraints \eqref{eq:ev5}, \eqref{eq:pw1} and \eqref{eq:pw2} allows a comparison to the current uni-directional charger costs as these new constraints stop the EVs from being able to discharge and provide upward balancing services (i.e.~frequency response services). For this scenario, introducing a flexible trip start time is comparable to moving the charging event itself. The only difference is the window of travel (i.e.~when the EVs may perform journeys) is more limited than the full 24 hours or available plug in times.

The prices used for summer and winter months are based on July 2019 and February 2020 wholesale electricity market prices in GB, respectively, in order to remove the distortion brought about by the Covid-19 pandemic. The timing of these prices was chosen to align with the months from which the travel data was collected, as described in Section \ref{fleet data}.

\begin{equation} \label{eq:ev10} 
    D_{t} = 0, \forall t \in T^{gr}
\end{equation}

\begin{equation}\label{eq:ev11} 
    BSup_{t}^{c} = BSup_{t}^{d} = 0, \forall t \in T^{gr}
\end{equation}

%%%%%%%%%%%%%%%%%%%% scenario 2 - smart charging charging %%%%%%%%%%%%%%%%%%%%%%

The second scenario is the smart charging scenario, where EVs charge at times where the cost of energy is lowest. The effect of moving the trip times in this case would only improve revenue if the original trips occurred when the price of energy is at the lowest point in the day. The modification of the original constraints is the same as for dumb charging, as shown in \eqref{eq:ev10} and \eqref{eq:ev11}.

%%%%%%%%%%%%%%%%%%%% scenario 3 - DC %%%%%%%%%%%%%%%%%%%%%%%%%%%%%%%%%%%%%%%%%%

The third scenario is the newest frequency response product on the GB market at the time of writing, Dynamic Containment \cite{NationalGridESO2021Markets2025}, which represents the current state of the market. The prices reflect the value of frequency response at present, in markets which are constantly changing and being updated. When the newest frequency product was launched, it was the first day-ahead procured response product but had a constant price throughout each day. The prices of this product in this scenario are therefore assumed to be constant \cite{NationalGridESODynamicContainment}, something that does not reflect the value of balancing the system and the time-varying requirement for frequency response products. This is because when system inertia is low, i.e.~renewable energy on the system is high and therefore net demand is low, the value of frequency response is significantly higher than in high-inertia conditions \cite{Badesa2020TowardsSystems}. Nonetheless, the prices seen in the Dynamic Containment market so far are higher than for previous fast frequency response products, such as Enhanced Frequency Response, which was procured through years-long contracts \cite{NationalGrid2016EnhancedQuestions}. Trends show that the prices of frequency response have increased over time.

%%%%%%%%%%%%%%%%%%%% scenario 4 - Future FR %%%%%%%%%%%%%%%%%%%%%%%%%%%%%%%%%%%%

The fourth and final scenario is the future frequency response scenario. Here, the price of frequency response is a reflection of the value of this service to a future system. The methodology by which these prices were generated is described in Section \ref{Luis prices} below.

\subsection{Modelling Future Frequency Response Prices in Great Britain}\label{Luis prices}
% Luis's description of how the price reflects the value

A view of future prices for the fast frequency response service in GB (referred to as Dynamic Containment, DC) is produced here to estimate the potential revenue from EVs with V2G capability. To produce this view, a frequency-constrained model is used, which outputs the projected price for DC for a given hour, based on the need for this service under each system condition (i.e.~level of demand and generation mix).

A demand curve with seasonal and daily trends was used, consistent with National Grid's projections for 2030 \cite{NationalGridESO2020FutureScenarios}, ranging from 20 GW to 60 GW of instantaneous hourly demand. A total of 45 GW of onshore and offshore wind combined was assumed, along with 35 GW of solar PV. A nuclear capacity of 5 GW was assumed, also in line with \cite{NationalGridESO2020FutureScenarios}. Given these system characteristics, the frequency response price for each hour of the future frequency response scenario / product is computed as follows:
\begin{enumerate}
    \item Net-demand is obtained by subtracting RES generation from national demand. The level of inertia in the system is estimated by considering that net-demand is covered by nuclear units (5GW with a 5s inertia constants) and Combined-Cycle Gas Turbines with an inertia constant of 5s.
    
    \item If the level of inertia obtained in Step 1 is below the minimum acceptable value for this magnitude, inertia is fixed at this minimum. This `floor' for inertia is determined by the Rate-of-Change-of-Frequency (RoCoF) limit, as defined by the expression below. Please refer to \cite{Badesa2020TowardsSystems} for full details on the impact of inertia on RoCoF.
    \begin{equation} \label{eq:RoCoF}
        \textrm{H}_\textrm{floor} = \frac{\textrm{P}\textsubscript{infeed}\cdot f_0}{2 \cdot \textrm{RoCoF}\textsubscript{limit}} 
    \end{equation}
    In \eqref{eq:RoCoF} $\textrm{H}_\textrm{floor}$ is the floor for inertia, $P\textsubscript{infeed}$ is the largest power infeed in the national system (considered to be 1.8GW), $f_0$ is the nominal frequency of the grid (50Hz in GB) and $\textrm{RoCoF}\textsubscript{limit}$ is the RoCoF limit (of 1Hz/s in the future GB system).
    
    \item Given the level of inertia obtained from Steps 1 and 2, the necessary volume of Primary Frequency Response (PFR) for this system condition is computed. Note that PFR refers to the slower frequency response provided by electromechanical devices, such as synchronous generators. To compute this necessary volume of PFR, the system condition for respecting that the frequency nadir will not cause the activation of Under-Frequency Load Shedding is used \cite{Badesa2019SimultaneousCommitment}:
    \begin{equation} \label{eq:nadir}
        \left(\frac{\textrm{H}}{f_0} - \frac{\textrm{DC}\cdot1\textrm{s}}{4\cdot \Delta f\textsubscript{max}}\right)\cdot \frac{\textrm{PFR}}{10\textrm{s}} \geq \frac{(\textrm{P}\textsubscript{infeed} - \textrm{DC})^2}{4\cdot \Delta f\textsubscript{max}}
    \end{equation}
    
    % decision variables are DC, H and PFR - how much extra DC can reduce the need for both H and PFR (they are linked) 
    % in CO2 section: how much H and PFR can be replaced with extra DC from the EVs (goes back to constraint 20)
    % we consider the nadir constraint 20 
    
    % for CCGT we assume 500MW plant with a 5s inertia constant operating at minimum stable generation (250MW) - this plant will be there just for ancillary service.. it has to provide energy in order to provide inertia. they limit how much RE we can have (on the generation side). 
    
    % we calculate for each hour of the month - the DC provided by the EVs then solved equation 20 subtracting the DC
    % (assumed equality is binding) 'vpasolve' in matlab, the unknown is x
    % take the average from hour to hour because20 is non linear so if we took averages for the month then solved only once we'd get a different answer

    where $\textrm{H}$ is the level of inertia, $\textrm{DC}$ is the volume of Dynamic Containment (required to be fully delivered in 1 second after a large contingency), $\Delta f\textsubscript{max}$ is the lowest nadir admissible to avoid Under-Frequency Load Shedding (of 0.8Hz in GB), and $\textrm{PFR}$ is the volume of PFR (delivered in 10 seconds by synchronous generators).
    
    The above condition guarantees that frequency stability will be maintained in the system, and therefore defines the requirement for ancillary services: it defines the required balance between volumes of inertia, size of the largest power infeed, and the two frequency response services, in order to avoid Under-Frequency Load Shedding. The interested reader is referred to \cite{Badesa2019SimultaneousCommitment} for full details on how this constraint is deduced.
    
    \item Given that the value of all magnitudes in eq.~(\ref{eq:nadir}) is defined, except that of PFR, this unknown can be directly solved. The volume of DC is considered to be of 1 GW (as per current plans in Great Britain \cite{Badesa2021AncillaryFuture}) and the size of the largest power infeed is of 1.8 GW (driven by a nuclear station expected to be commissioned in GB in coming years).

    \item Finally, the price of DC is computed from the volume of PFR that an additional MW of DC could replace, obtained from the balance described by eq.~(\ref{eq:nadir}). A price of \pounds10/MW/h for PFR was considered, in line with average prices seen for this service in GB in recent years \cite{NationalGridESO2020Firm2020}.

\end{enumerate}

A sensitivity on these projected prices is considered in Section~\ref{SensitivityPrice}, to provide a comprehensive view on V2G revenues under different scenarios for future prices for Dynamic Containment.

\section{Case Studies}{\label{case studies}}

In this section, we present the inputs and the results for several case studies chosen in a way as to be closest to real world future scenario and to demonstrate the value of the approach. Section~\ref{fleet data} describes the input fleet data, including trends of driving patterns of two fleet types. The approach to generate synthetic data based on these trends is also included. Section~\ref{results and discussion} presents the results from the various operational scenarios described in Section~\ref{scenarios and prices}. Then, the technique applied to reduce battery cycling in the modelling is described in Section~\ref{battery cycling}. Results of sensitivity analysis, key to understanding how susceptible the revenues derived are to changes in frequency response price and traffic conditions, are shown in Section~\ref{SensitivityPrice} and Section~\ref{SensitivityTraffic}, respectively. Finally, the consequence of the given approach on \text{$CO_2$} emissions is quantified in Section~\ref{CO2 avoided}.

\subsection{Fleet Data}{\label{fleet data}}
Derived from the Innovate UK funded project E-Flex \cite{ProjectE-Flex}, the driving patterns from two commercial fleets participating in the trial were analysed and synthetic data was generated. Two types of commercial fleets were considered: i) a delivery fleet with 10 EVs, ii) a maintenance fleet with 5 EVs. The reason that these fleets were chosen in that commercial fleets have regular, defined trips that are scheduled during weekdays. Compared to private vehicles, commercial fleets are well suited to balancing service provision with V2G due to their coordination by fleet operators, making them easier to aggregate. The main differences between the two fleets are summarised in Table~\ref{tab:fleets} for winter and summer months. For the delivery fleet, although the mean travel duration and trip length range increase in the summer compared to the winter, the mean travel energy decreases. The opposite is true for the maintenance fleet, which has a higher mean travel energy in the summer compared to winter, but a shorter mean travel duration. The difference between winter and summer for the maintenance fleet and its implications is discussed in Section \ref{months RBG only}. The differences between the fleet types are examined in Section \ref{summer both fleets}.
\begin{table}[h]
\caption{The characteristics of the two commercial fleets in winter and summer months.}
\centering
\begin{tabularx}{1\columnwidth}{c| *{4}{c}}
\hline
\textbf{} & \multicolumn{2}{c}{\textbf{Winter}}
& \multicolumn{2}{c}{\textbf{Summer}} \\ & Delivery & Maintenance & Delivery & Maintenance \\
\hline
Travel start time mean & $10.30\text{am}$ & $8.00\text{am}$ & $7.30\text{am}$ & $8.00\text{am}$ \\ \hline
Charging start time mean  & $2.30\text{pm}$ & $3.00\text{pm}$ & $1.30\text{pm}$ & $12.00\text{pm}$ \\ \hline
Travel duration mean & $5\text{h}$ & $7.5\text{h}$ & $6.5\text{h}$ & $6.5\text{h}$ \\ \hline
Travel duration min \& max & $0.5\text{h}-6\text{h}$ & $7\text{h}-8\text{h}$ & $0.5\text{h}-9\text{h}$ & $5\text{h}-8\text{h}$ \\ \hline
Travel energy mean & $8.9\text{kWh}$ & $4.2\text{kWh}$ & $7.7\text{kWh}$ & $5.7\text{kWh}$ \\ \hline
\end{tabularx}
\label{tab:fleets}
\end{table}

\begin{figure}[h]
\centering
\includegraphics[width=0.6\textwidth, trim =0 0 0 0, ]{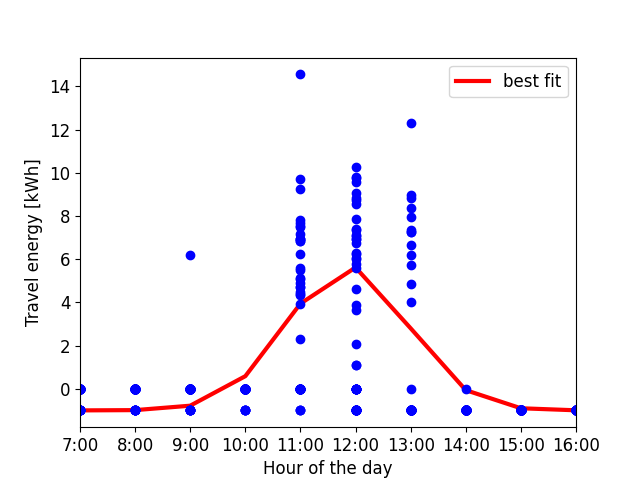}
\caption{The driving patterns distribution for the month of July of the maintenance fleet and the optimised fitting function in red.}
\label{fig:gaussian}
\end{figure}

The 24-hour driving patterns $x(t)$ of the commercial fleets showed $x(t)=-1$ for time steps $t$ when the EV was connected to the grid, $x(t)=0$ for time steps $t$ when the EV was on a trip and the energy needed for the travel at the first time step $t_{return}$ the EV was connected again to the grid following the travel. Values of $x(t)<0$ do not indicate any energy exchanges, e.g.~energy feeding into the grid, but only that the EV is not on a trip. Two months were chosen to be representative of winter and summer months, i.e.~February and July respectively. For each of these two months, the driving patterns for each of the two commercial fleets were fitted with the Gaussian function defined in Equation~\eqref{eq:gaussian}~\cite{Vagropoulos2013},~\cite{Xu2017}.
\begin{equation}\label{eq:gaussian} 
f(x(t))=a\cdot\exp^{-\cfrac{(x(t)-b)^2}{c^2}} \ \ \ \  \forall t\in[0,T]
\end{equation}
In Equation~\eqref{eq:gaussian} $a$ is the height of the curve's peak, $b$ is the position of the center of the peak, and $c$ is the Gaussian width. These parameters were then optimised to improve the fitness of the function to the real patterns using non-linear least squares analysis, resulting in residual standard deviations less than one. The fitting function for the month of July of the maintenance fleet is shown in Figure~\ref{fig:gaussian}. The height of the curve's peak is relatively low as most of the hours the EVs were connected to the grid or on a trip. Once the optimal fitting curve was obtained, the parameters $a\in[0,\text{max}\{x(t)\}]$ and $b\in \{t|f(x(t))>0\}$ were randomly sampled to generate synthetic driving patterns. The parameter $c$ was not varied as it defines the length of the trip that is distinctive for each fleet. The $24$-hour driving patterns of $100$ EVs for each of the two fleets were finally generated for the months of February and July.

\subsection{Results and Discussion}\label{results and discussion}

Following the details of the case studies set out in Section \ref{case studies}, the results from the case studies are presented and findings are discussed here. . The results make a comparison between both the summer and winter seasons, showing the effects of renewable generation on ancillary services prices and hence revenue. The revenues for the inflexible (i.e.~original) trips and the flexible trips across the summer and winter seasons were first compared using the maintenance fleet to model the four scenarios outlined in Section~\ref{scenarios and prices}. Then, the difference between the maintenance and delivery fleets was investigated to show how driving patterns affect revenue in this approach. Lastly, some extreme weather days were analysed to show examples of potential maximum benefit from flexible trips. This is a way to demonstrate the major differences between fleets and seasons in an extreme case.

The optimisation problem defined in Section \ref{methodology} was modelled in Python using the Gurobi Python Environment, with a MIP gap of 0.1\%. It was run on a machine with a 3.6GHz Intel Core i7-7820X 8-Core processor and 64 GB of RAM. The average computing time was 20 minutes.

\subsubsection{Optimal Start Times in Different Months for Maintenance Fleet}\label{months RBG only}

\begin{figure}[t]
    \includegraphics[width=\linewidth]{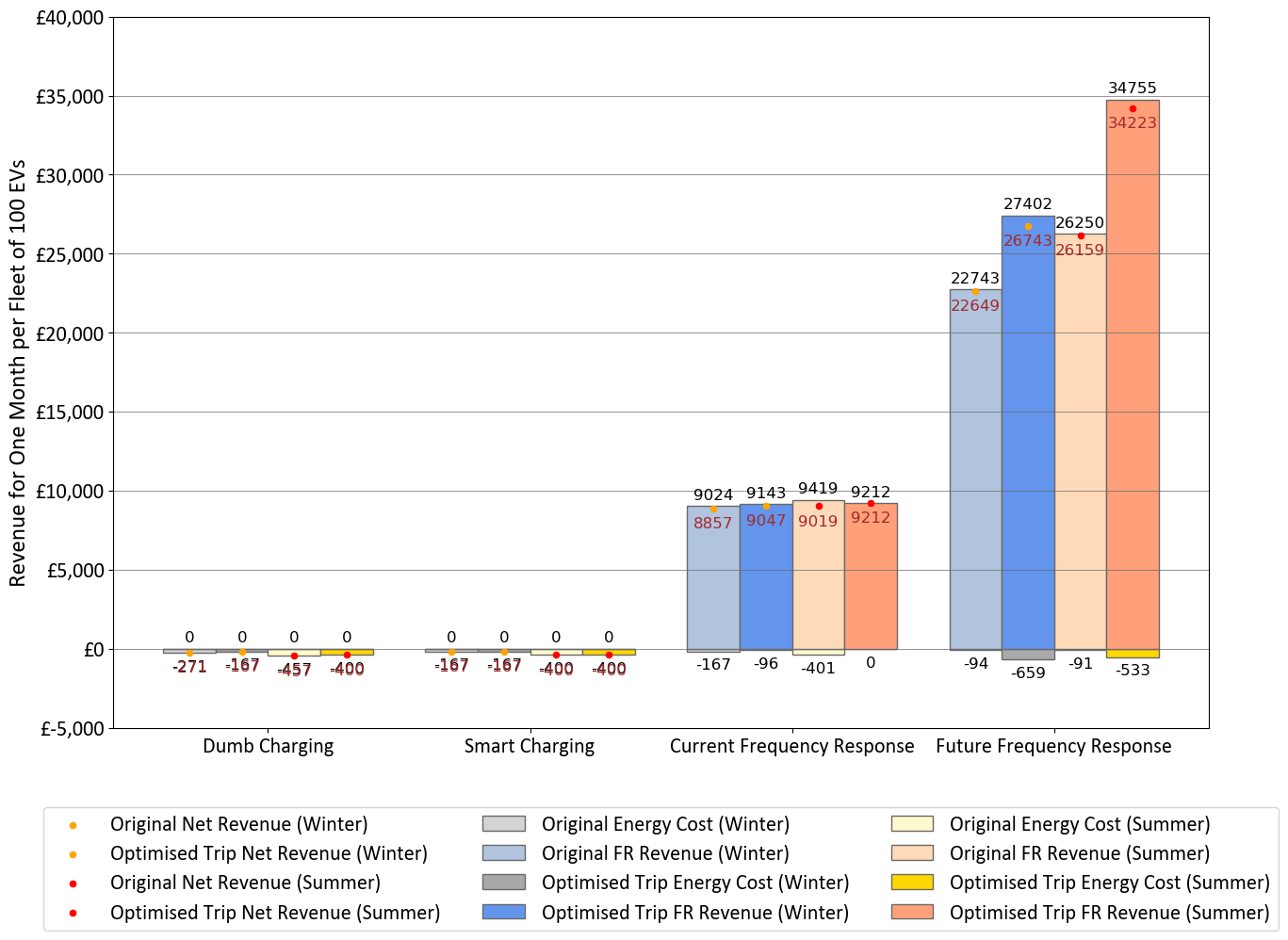}
    \caption{Summer and Winter Months for Maintenance Fleet with and without Optimal Trip Scheduling}
    \label{fig:fig1}
\end{figure}

Using the fleet data described in Section \ref{fleet data}, the four scenarios were simulated for both the summer and winter months with the original start time of the trips, then with optimal trips. The maintenance fleet results are shown in Figure \ref{fig:fig1}. The dumb charging scenario of the maintenance fleet shows that the charging costs (i.e.~energy costs for travel) are higher in the summer, due to further distances travelled (see Table~\ref{tab:fleets}) and higher electricity prices (as described in Section \ref{scenarios and prices}). By making the timing of the journeys flexible, there is a maximum cost saving of 38\% in the winter month for this scenario, relative to the original travel pattern, compared to the cost saving of 12\% in the summer. This means when charging on arrival in the winter, there is relatively more benefit in moving this plug in time (by moving the trip) in order to capture lower energy prices and reduce costs. Averaging this benefit over a whole year, a single EV from the maintenance fleet can make an additional annual revenue of up to £729 by incorporating flexible journeys into its operations.

Changing the trip times in the smart charging scenario does not save on costs because the charging will occur at night in either case, when the prices are lowest and when the EVs are connected.

By adding bidirectional (V2G) charging, the EVs are now able to participate in the frequency response market, resulting in positive net revenues for the third and fourth scenarios. In the third scenario, having flexible start times does not generate a significant increase in revenue, due to the fact that current frequency response prices used are constant for every hour in the day. 

By avoiding traveling in times of high future frequency response prices, optimal trip scheduling allows for more revenue from this service in both months of the fourth scenario. The energy costs also reduce with optimal trips, due to the flexibility of the start time, where batteries must be charged to the required level. An assumption is made that when scheduling flexible trips, the 24-hour period begins with a 50\% state of charge (or the maximum state of charge in order for the EV to depart with the 95\% requirement). 

It is also worth noting that the energy costs increase in the fourth scenario due to the fact that it is beneficial to charge at time periods of high frequency response prices. When charging, the EV will get paid for offering availability to reduce its charging and increase its discharging, i.e.~as per objective function \eqref{eq:obj}. As the EVs are plugged in when frequency response are higher, they will charge at these times and although they will be charged for the energy, it still results in a net revenue that is higher than the original trip times.

\begin{figure}
    \centering
    \includegraphics[width=\linewidth]{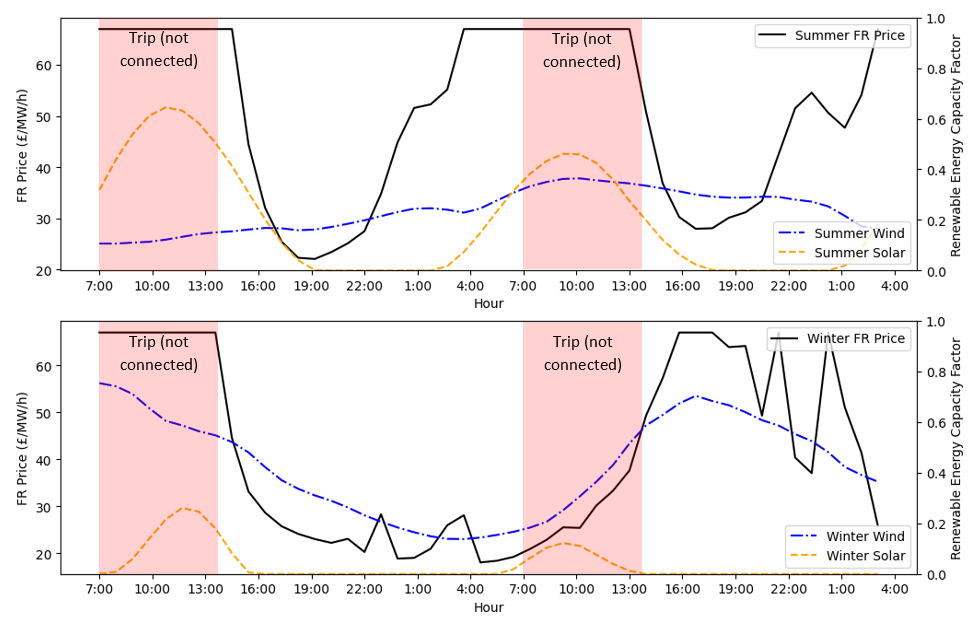}
    \caption{Example Summer and Winter Days Comparison}
    \label{fig:prices}
\end{figure}

In the future frequency response scenario, fleet revenues are expected to rise by 156\% in the winter and 190\% in the summer with original trips or driving patterns, as compared to the current frequency response scenario. When trip start times are optimised in this scenario, the revenues increase by 18\% in the winter and 31\% in the summer, as compared to the same scenario with original trip times. As described in Section~\ref{Luis prices}, the frequency response prices depend on net demand, which in turn depends on renewable energy generation. The reason why introducing optimal start times results in a larger improvement in the summer month is that solar output occurs during the day, driving up the price of frequency response. The maintenance fleets are able to use this to their benefit and schedule their trips for later in the day, in order to maximise revenue by remaining plugged in at these times in order to provide frequency response services. In the winter, the dominant renewable generation is wind, which is usually strongest at night, when fleets are not typically operational. An example of this seasonal difference is shown in Figure~\ref{fig:prices}, where two days of frequency response prices show a regular pattern in the summer month. The original trip times occur during peak frequency response prices and benefit from being moved to later in the day, when frequency response prices are lower. In the winter, however, wind and hence prices do not follow such a regular pattern. Although during the first day of winter in this example, it is beneficial to move the trip to later in the day, for the second winter day it is best to travel as originally scheduled.

\subsubsection{Optimal Start Times in Summer Month for Maintenance and Delivery Fleets} \label{summer both fleets}

\begin{figure}
    \includegraphics[width=\linewidth]{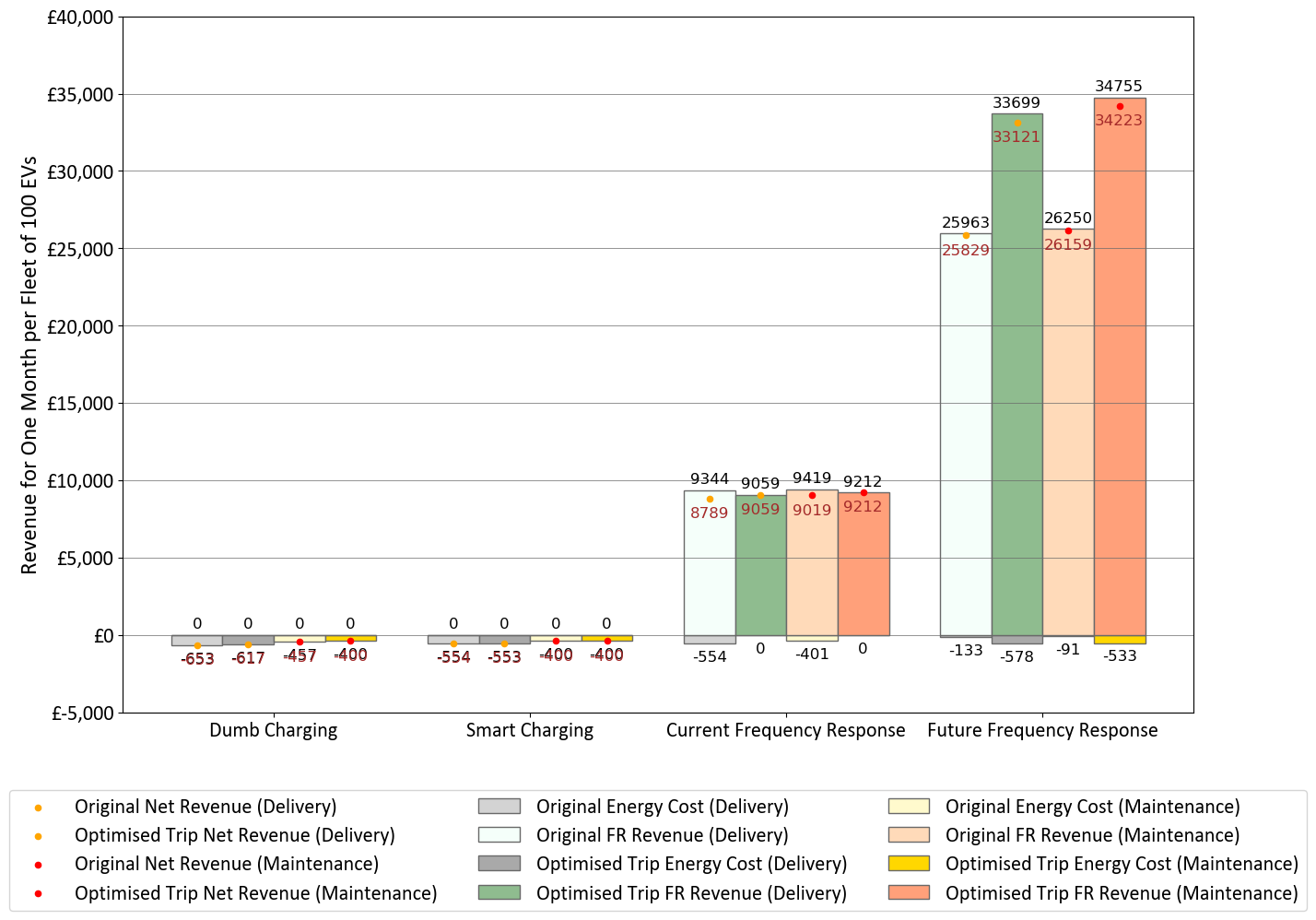}
    \caption{Comparison of Delivery and Maintenance Fleets in Summer Month with and without Optimal Trip Scheduling}
    \label{fig:fig2}
\end{figure}

In Figure \ref{fig:fig2}, the delivery fleet data is introduced in comparison to the maintenance fleet in the summer month, in order to understand the effect of fleet type on revenues. As shown in Table \ref{tab:fleets}, the mean travel duration is the same for these fleets, but the delivery fleet will travel further, using more energy. Also, the maintenance fleet makes more consistent journeys, all between five and eight hours long, whereas the delivery fleet has a much shorter minimum journey length and a longer maximum journey length. In addition to a different travel pattern, the windows in which the fleets are assumed to travel within are different. With flexible start times, the maintenance fleet is able to travel in a 22 hour window, so from 7am to 5am, allowing time to return to base and charge before the next day. When planning fleet logistics, the delivery fleet will likely have more limitations in when its trips can be made, and so the travel window is set to 16 hours, between 7am and 11pm.

 In the current frequency response scenario, revenues from optimising trip times are similar compared to the original trips, as frequency response prices are static throughout the 24-hour period. The maintenance fleet has slightly more revenue, as the trips have a lower mean travel energy. However, in the future frequency response scenario, the maintenance fleet has a higher increase in revenue from flexible journeys of 31\% compared to the delivery fleet of 28\%. This is because the original trips made by the maintenance fleet were consistently longer than the delivery fleet, so a greater percentage of the trips would benefit from having a flexible start time, capturing revenues from providing frequency response by shifting the journey later in the day.

\subsubsection{Optimal Start Times on Future Extreme Weather Days for Maintenance and Delivery Fleets}\label{extreme days}

\begin{figure}
    \includegraphics[width=\linewidth]{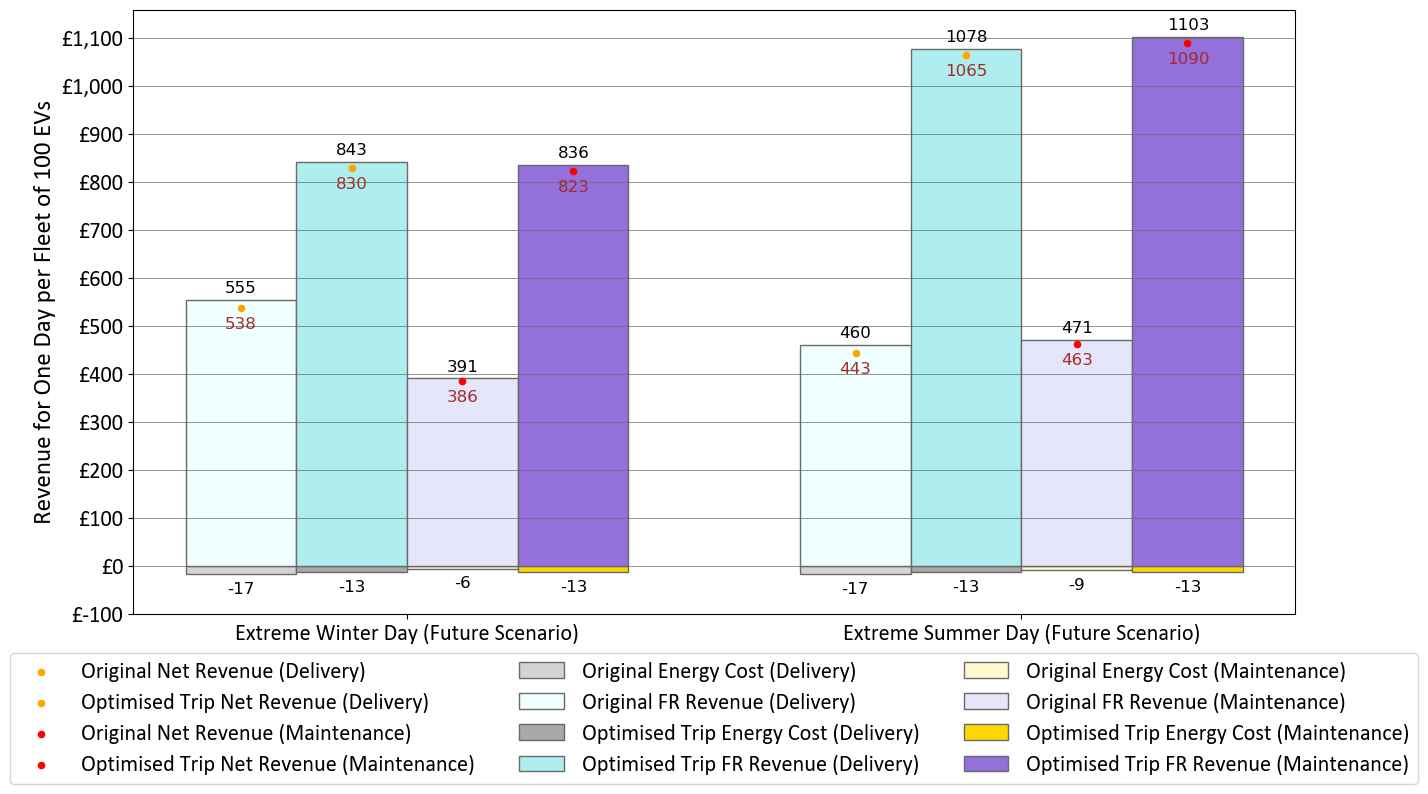}
    \caption{Extreme Summer and Winter Days for Delivery and Maintenance Fleets with and without Optimal Trip Scheduling in Future Scenario}
    \label{fig:fig3}
\end{figure}

Assuming that in the future, prices of frequency response products will reflect net demand and hence system inertia, some extreme weather days (i.e.~days with the most favourable weather conditions for increasing revenues through flexible journeys) have been simulated to investigate the maximum potential benefit of flexible trip scheduling. In Figure \ref{fig:fig3}, revenues during an extreme winter and summer day are shown for both fleets. The summer scenario demonstrates more benefit from optimal trip scheduling, with the maintenance fleet and delivery fleet increasing their revenues by 135\% and 140\%, respectively, compared to original trip schedules. This is due to how valuable frequency response will be during daylight hours on days where solar generation is high and net demand is low.

\begin{figure}
    \includegraphics[width=0.8\linewidth]{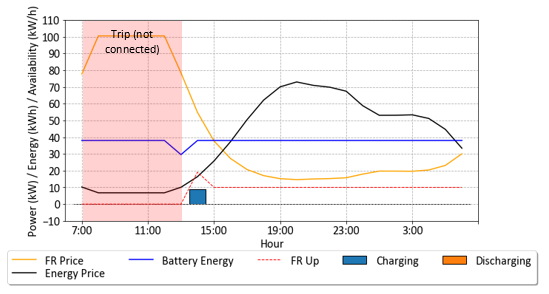}
    \centering
    \caption{Extreme Summer Day: Original Trip by Maintenance Fleet Vehicle}
    \label{fig:extreme_original}
\end{figure}

\begin{figure}
    \includegraphics[width=0.8\linewidth]{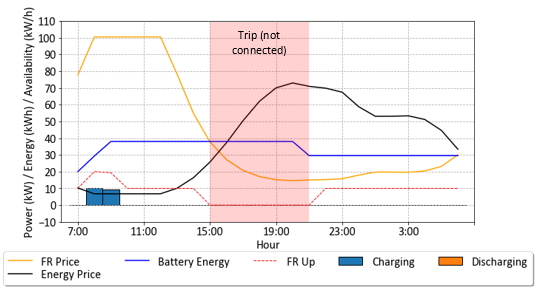}
    \centering
    \caption{Extreme Summer Day: Optimal Trip by Maintenance Fleet Vehicle}
    \label{fig:extreme_opt_trip}
\end{figure}

Figure \ref{fig:extreme_original} shows an example original trip schedule over a 24-hour period for a maintenance fleet vehicle on an extreme summer day. As the vehicle is on a trip from 7am to 1pm, it is not able to offer frequency response when the prices are high. In Figure \ref{fig:extreme_opt_trip}, the trip has been optimised and starts in the afternoon, when frequency response prices are lowest. This way, the EV can provide maximum frequency response earlier in the day by staying plugged in and delaying its journey.

\subsection{Method to Reducing Battery Cycling}\label{battery cycling}

\begin{figure}
    \includegraphics[width=0.7\linewidth]{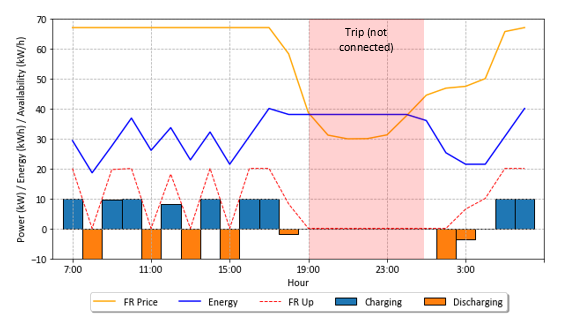}
    \centering
    \caption{Summer Day: Optimal Trip by Maintenance Fleet Vehicle}
    \label{fig:nopen}
\end{figure}

When prices for frequency response are high, the optimal charging behaviour of the EVs involves `cycling' (i.e.~charging then discharging) in order to increase revenues, as shown in Figure \ref{fig:nopen}. This is profitable for the EV if charging and discharging efficiencies are considered, otherwise providing two time periods of charging fully and then discharging would be equal to two time periods of neither charging or discharging. Due to losses associated with efficiencies, energy is dissipated, allowing for the EV battery to commit to more in availability than it acquired in energy. This effect leads to repeated cycles of battery charging-discharging to increase frequency response revenues. 

\begin{figure}
    \includegraphics[width=0.7\linewidth]{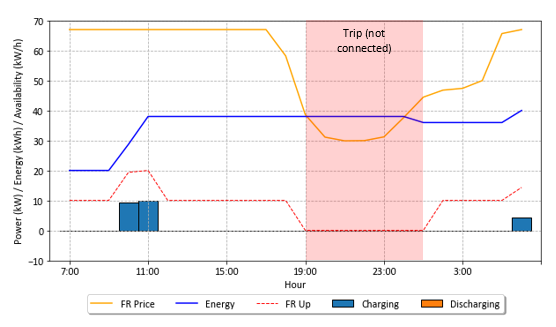}
    \centering
    \caption{Summer Day: Optimal Trip by Maintenance Fleet Vehicle with Penalty Applied}
    \label{fig:pen}
\end{figure}

As this unintended effect has a detrimental impact on battery health \cite{Bishop2013EvaluatingEV}, a penalty term ($\delta$) was applied to the discharging term ($D_t$) as per equation~\eqref{eq:pen}. By applying a penalty to discharging, this minimises battery cycling, while only minimally affecting revenues from providing frequency response availability. This is shown in Figure~\ref{fig:pen} for the same summer scenario. This reduces revenues in this case by 3\%.

\begin{equation} \label{eq:pen} 
    \text{maximise} \left (\sum_{t\epsilon T^{grid}}  (BSup_{t}^{c}+BSup_{t}^{d}) \cdot \pi_t^{BSup} - C_{t} \cdot \pi_t^{b} - D_{t} \cdot \pi_t^{s} \cdot \delta \right)
\end{equation}

The appropriate value for the penalty `$\delta$' depends on the maximum frequency response prices than may be captured by the EV: the penalty must be sufficiently high as to avoid the revenue gains from cycling. In this paper, the value used was $\delta=0.5$.

\subsection{Sensitivity Analysis: Frequency Response Prices} \label{SensitivityPrice}

\begin{figure}
    \includegraphics[width=\linewidth]{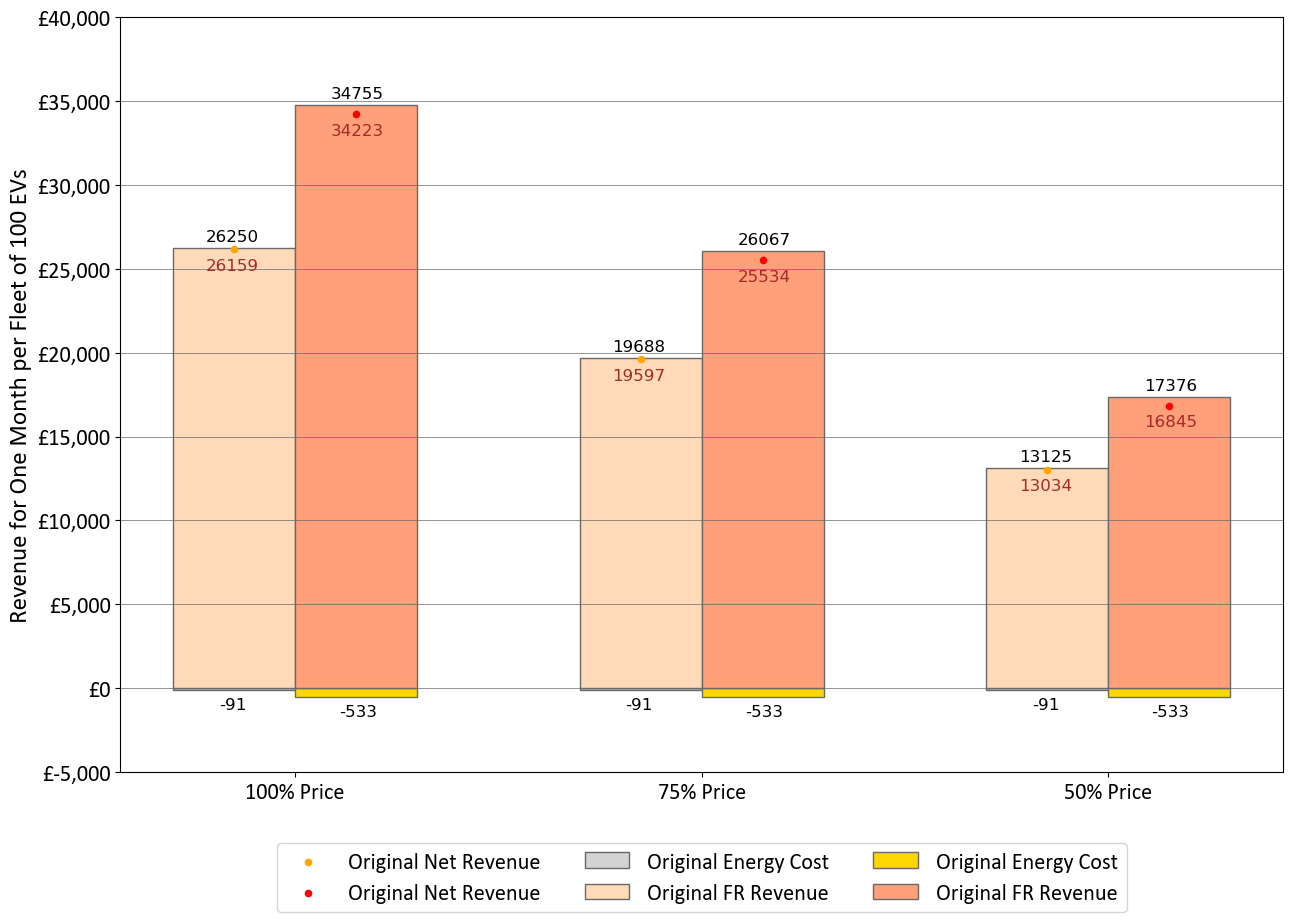}
    \caption{Sensitivity Analysis of Maintenance Fleet Summer Revenues under 3 DC Price Scenarios}
    \label{fig:sens}
\end{figure}

In order to demonstrate how sensitive the revenues for these fleets were to future frequency response prices (see Section~\ref{Luis prices}), sensitivity analysis was performed for three scenarios, as shown in Figure~\ref{fig:sens}. The way this analysis was performed was to run the original optimisation, bench-marked against the same prices but no optimal trips (as in Section~\ref{results and discussion}, first with the full future frequency response price (100\%). The frequency response price was then reduced to 75\% and then 50\%, in order to see the impact on revenues that a reduced price would have. This is a way to investigate the uncertainty of these prices. 

Based on the results of the sensitivity analysis, the value of flexible trips compared to original trip schedules is consistently around 23\%. Meaning, the additional revenue gained from scheduling trips is proportionally the same, regardless of the frequency response prices. In other words, the price does not change how much added value the approach brings. A 25\% reduction in frequency response price led to a similar subsequent reduction in revenue (for both original and optimal trip schedules) of 25\%. However, a reduction of the frequency response price by another \%25 led to a greater reduction in revenues of 33\% - 34\%, relative to the 75\% case. 

\subsection{Sensitivity Analysis: Traffic-Related Factors} \label{SensitivityTraffic}

It is recognised that there is a degree of uncertainty around trip length. It could be that trip length is affected by external factors including road blockages and traffic conditions. This uncertainty in trip length is explored by performing a sensitivity analysis to understand the impact of such factors on the revenue and hence the usefulness of the approach, particularly in areas where external factors are likely to impact trip length. These include cities, for example, where commercial fleets are likely to operate.

\begin{figure}
    \includegraphics[width=\linewidth]{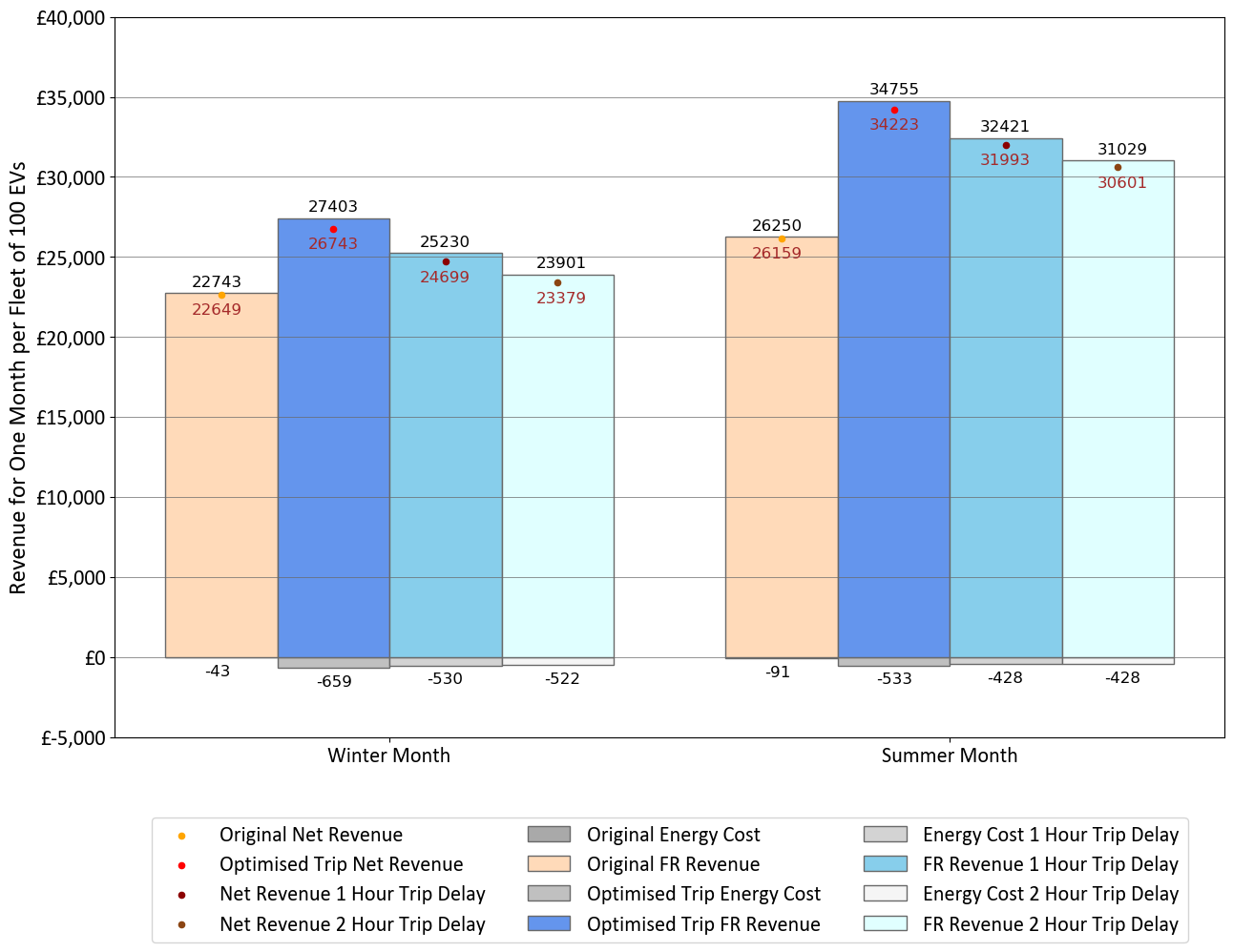}
    \caption{Sensitivity Analysis of Maintenance Fleet Revenues Considering Trip Length Delays}
    \label{fig:sens_trips}
\end{figure}

As shown in Figure~\ref{fig:sens_trips}, it was found that for the maintenance fleet, maximum revenues attainable with optimal trips in the summer were reduced by 7\% by extending the trip length by one hour and by 11\% by extending the trip length by two hours. For the same fleet, in the winter, maximum revenues were reduced by 8\% by extending the trip length by one hour and by 13\% by extending the trip length by two hours.

The reason the EV fleet loses more revenue as a result of a longer trip delay in the winter is similar to the reasons less revenue was added by flexible trip times in Section~\ref{months RBG only}. During winter, there is more price variability and less regular daily weather patterns dominated by solar generation (as in the summer). If the trip were to be delayed to the evening on some days where wind was increasing as the day went on and FR prices were increasingly, then the EV fleet revenue will be more susceptible to being impacted by longer trips (i.e. trip delays).

\subsection{$CO_2$ Emissions Avoided}\label{CO2 avoided}

By providing frequency response, the fleets of EVs can avoid resorting to Combined Cycle Gas Turbines (CCGT's) standing by to respond to grid contingencies, therefore emitting greenhouse gases. Here we compute the system savings in terms of carbon emissions due to the frequency response provided by EVs, assuming an emission rate of $368$ g$CO_2$ equivalent/kWh for CCGTs. The results are presented in Table~\ref{tab:emissions} for a larger fleet of 5,000 EVs with the same characteristics as the ones used in Section~\ref{case studies}.

\begin{table}[h] 
\caption{CCGTs and kg of $CO_2$ avoided per month per 5,000 EVs}
\centering
\begin{tabularx}{0.9\columnwidth}{c| *{4}{c}}
\hline
\textbf{} & \multicolumn{2}{c}{\textbf{Winter}}
& \multicolumn{2}{c}{\textbf{Summer}} \\ & Delivery & Maintenance & Delivery & Maintenance \\
\hline
CCGTs Avoided & 1.1 & 1.15 & 1 & 1.05 \\ \hline
$CO_2$ Avoided (kg) & 67,000 & 71,000 & 62,500 & 66,500 \\ \hline

\end{tabularx}
\label{tab:emissions}
\end{table}

The number of CCGTs avoided was calculated using the nadir constraint \eqref{eq:nadir} described in Section~\ref{Luis prices}, solving for $x$ as per equation \eqref{eq:nadirCO2}. The variables in the nadir constraint \eqref{eq:nadir} are $\textrm{H}$, $\textrm{PFR}$ and $\textrm{DC}$. As $\textrm{H}$ and $\textrm{PFR}$ are both provided simultaneously by synchronous generators (e.g.~CCGTs), it is possible to input the contribution in $\textrm{DC}$ from the EVs (i.e.~$\textrm{DC}\textsubscript{EVs}$) and see what the subsequent reduction in provision of $\textrm{H}$ and $\textrm{PFR}$ is from the CCGTs (where $x$ is the number of CCGTs avoided). It is assumed that each CCGT is a 500MW plant with a 5s inertia constant, operating at a minimum stable generation of 250MW. The CCGTs are able to provide 75MW (i.e.~15\% of their capacity) for PFR. As the equation is non-linear, the calculation was solved hour by hour and then averaged for the month to produce the results in Table~\ref{tab:emissions}.

\begin{equation} \label{eq:nadirCO2}
\begin{split}
    \left[ \frac{\textrm{H}\textsubscript{tot}+x\cdot \textrm{H}\textsubscript{CCGTs}}{f_0} - \frac{(\textrm{DC}\textsubscript{tot} - \textrm{DC}\textsubscript{EVs})\cdot1\textrm{s}}{4\cdot \Delta f\textsubscript{max}}\right]\cdot & \frac{\textrm{PFR}\textsubscript{tot} + x\cdot \textrm{PFR}\textsubscript{CCGTs}}{10\textrm{s}} = \\
    & \frac{[\textrm{P}\textsubscript{infeed} - (\textrm{DC}\textsubscript{tot}-\textrm{DC}\textsubscript{EVs})]^2}{4\cdot \Delta f\textsubscript{max}}
\end{split}
\end{equation}

    % decision variables are DC, H and PFR - how much extra DC can reduce the need for both H and PFR (they are linked) 
    % in CO2 section: how much H and PFR can be replaced with extra DC from the EVs (goes back to constraint 20)
    % we consider the nadir constraint 20 
    
    % for CCGT we assume 500MW plant with a 5s inertia constant operating at minimum stable generation (250MW) - this plant will be there just for ancillary service.. it has to provide energy in order to provide inertia. they limit how much RE we can have (on the generation side). 
    
    % we calculate for each hour of the month - the DC provided by the EVs then solved equation 20 subtracting the DC
    % (assumed equality is binding) 'vpasolve' in matlab, the unknown is x
    % take the average from hour to hour because20 is non linear so if we took averages for the month then solved only once we'd get a different answer

%'complementary patterns' between commercial fleets and FR.. which will come out in the case study results. At night the EVs are all plugged in and done for the day so can participate in FR when wind is blowing and inertia is low. The opportunity for synchronicity between the commercial fleets delaying their journeys and providing FR is high. The control-ability of the commercial fleets is strong, the fleet being able to participate in the FR market (compared to private fleet etc) - this goes back to one of the reasons commercial fleets and FR make up the chosen scenario for this.
\section{Conclusion and Future Work}\label{conclusion}

This work has shown the significant potential benefits of EVs providing frequency response services in the current and future GB system scenarios, where flexibility is projected to be more valuable. Real-world data was used to accurately characterise the flexibility of commercial fleets using a methodology for scheduling optimal journey times according to frequency response prices. It was shown that such fleets with flexible journeys participating in these markets would increase their revenues by up to 38\% in the summer and 12\% in the winter. Fleets of an increased volume of 5,000 EVs with the same characteristics will be able to replace at least one CCGT plant for frequency control purposes, and the equivalent of at least 62,500 kg of $CO_2$ per month.

Given the demonstrated value of optimal trip scheduling for EV fleets, future work could develop this value quantification into an operational tool for fleet operators, considering uncertainties in driving patterns due to external factors such as traffic conditions, as well as renewable energy output. This will allow for minor adjustments to the schedule in real-time and considering changing conditions. Also, a wider range of fleet types and use cases, such as electric taxis or car-sharing fleets, could be considered. Certain fleets will prove more suited for this scheduling approach, and it will be important to identify them. Assumptions around fleet operations (e.g.~travel windows) could be further investigated and refined, to test whether a wide range of commercial fleets could benefit. Other ancillary services could be considered, including synthetic inertia provision and reactive power provision, which could potentially enhance the business case for flexible EV fleets as well as for other demand-side response assets. We would also aim to investigate techniques to incorporate a larger number of vehicles without compromising on the speed of reaching an optimal solution for trip scheduling. Furthermore, integrating non-commercial vehicles into the formulation, such as domestic EVs that are unlikely to change travel times but could provide frequency services when plugged in, would provide a comprehensive view of the potential system benefits of an electrified transportation sector.

\section{Acknowledgements}
This work was partially supported by the UK Engineering and Physical Sciences Research Council (EP/L015471/1) and by the Innovate UK project `E-Flex' (104249).

\bibliographystyle{elsarticle-num.bst}
\bibliography{references.bib}
\end{document}